\documentclass[twocolumn,superscriptaddress,preprintnumbers,amsmath,amssymb,pre]{revtex4-1}
\usepackage{graphicx}
\usepackage{dcolumn,xcolor}
\usepackage{amsmath} 
\usepackage{amssymb}
\usepackage{amsfonts}
\usepackage{bm}
\usepackage[latin1]{inputenc}
\usepackage{hyperref}

\begin{document}

%\widowpenalty=1000
%\clubpenalty=1000
%\preprint{APS/123-QED}

 \title{Drying of agarose gels monitored by \textit{in-situ} interferometry}

\author{Bosi Mao}
\email{mao@crpp-bordeaux.cnrs.fr}
\affiliation{Centre de Recherche Paul Pascal, CNRS UPR 8641 - 115 avenue Dr. Schweitzer, 33600 Pessac, France}
\author{Thibaut Divoux}
\email{divoux@crpp-bordeaux.cnrs.fr}
\affiliation{Centre de Recherche Paul Pascal, CNRS UPR 8641 - 115 avenue Dr. Schweitzer, 33600 Pessac, France}
\affiliation{MultiScale Material Science for Energy and Environment, UMI 3466, CNRS-MIT, 77 Massachusetts Avenue, Cambridge, Massachusetts 02139, USA}
\author{Patrick Snabre}
\email{snabre@crpp-bordeaux.cnrs.fr}
\affiliation{Centre de Recherche Paul Pascal, CNRS UPR 8641 - 115 avenue Dr. Schweitzer, 33600 Pessac, France}

\date{\today}

\begin{abstract}
Hydrogels behave as viscoelastic soft solids and display a porous microstructure filled with water with typical amounts of 90\%~w/w or more. As such, these materials are highly sensitive to water loss through evaporation, which impacts their mechanical properties. Yet, aside from scattered empirical observations, little is known about the gel drying kinetics for which there is a lack of temporally and spatially resolved measurements. Here we report a benchmark study of the slow drying of agarose gels cast in cylindrical Petri dishes. The weak adhesion of the gel to the lateral wall of the dish guarantees that the gel diameter remains constant during the drying process and that the gel shrinkage is purely vertical. The thinning rate is monitored by in-situ interferometry using a Michelson interferometer. The displacement of interference fringes are analyzed using an original spatiotemporal filtering method, which allows us to measure local thinning rates of about 10~nm/s with high accuracy. Experiments conducted at different positions along the gel radius show that the gel thins locally with a constant velocity before experiencing a sudden collapse at the end of the drying process. We further use the thinning rate measured at the center of the dish during the early stage of the drying process as a robust observable to quantify the role of additives on the gel drying kinetics and compare gels of different compositions. Our work exemplifies interferometry as a powerful tool to quantify the impact of minute amounts of additives on the drying of biopolymer gels.
\end{abstract}

%\pacs{83.50.Rp, 83.80.Hj, 83.60.La, 47.57.E-}
\maketitle
\section{Introduction}

  Hydrogels consist in a wide variety of soft viscoelastic solids that are commonly encountered in nature, as exemplified by hagfish slime \cite{Bocker:2016}, and lie at the core of numerous industrial applications such as scaffolds for tissue engineering  \cite{Vlierberghe:2011}, growth culture media \cite{Smith:2012,Jung:2016}, controlled drug release \cite{Rinaudo:2008,Hennink:2012}, etc. Hydrogels are mainly composed of water and contain only a few percent in mass of natural or synthetic polymers that are linked together either by covalent bonds, or by physical interactions such as hydrogen bonds, or dipole-dipole interactions, etc. \cite{Nijenhuis:1997,Djabourov:2013}. In both cases, polymers form a fibrous-like, sample-spanning network that is responsible for the gel viscoelastic behavior under low external strain \cite{Larson:1999}. Moreover, due to their fibrous structure composed of interconnected nonlinear springs, hydrogels experience a pronounced hardening under larger deformations \cite{Storm:2005,Pouzot:2006,Carrillo:2013,Mao:2016a}, up to the formation of macroscopic fractures, which are characteristic of a brittle rupture scenario \cite{Bonn:1998b,Leocmach:2014}.

%%%%%%%%%%%
\begin{figure*}[!t]
\centering
	\includegraphics[width=0.9\linewidth]{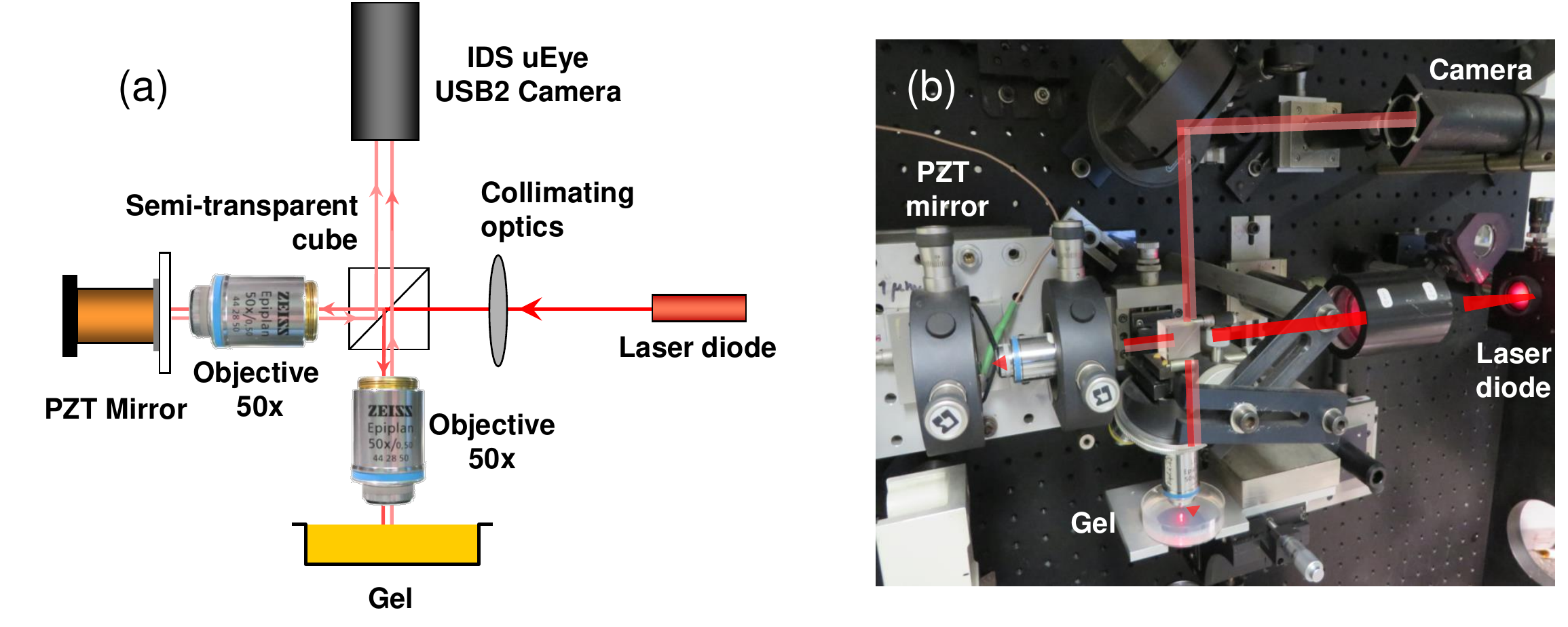}
\caption{(a) Sketch of the experimental setup that consists in a Michelson Interferometer operated in reflection mode. (b) Photo of the real experimental setup. The scale is set by the diameter of the Petri dish of 50~mm. The experimental setup is placed in a box that can be closed with thick curtains to limit air convection that may artificially increase the gel thinning rate.     
\label{fig1}}
\end{figure*} 
%%%%%%%%%%%

Being mainly composed of water, biopolymer gels are highly sensitive to water loss through evaporation and stress-induced solvent release \cite{Brown:2009a}. The drying of polymeric gels has been quantified by macroscopic observations \cite{Zrinyi:1993}, weighing \cite{Iglesias:1993}, and more local investigation techniques such as small angle neutron scattering \cite{Bastide:1984}, fluorescence spectroscopy \cite{Tari:2008}, and interferometry \cite{Wu:1994,Zhou:1996}. The goal of previous studies was to set the basis of a thermodynamics of swelling and shrinking, and among other things to test Li and Tanaka predictions for the drying kinetics of crosslinked polymeric networks \cite{Li:1990}. Previous experiments mainly consisted in monitoring the shrinkage of a disc-shaped gel with free boundary conditions. In the present work, we tackle the case of hydrogels cast in a cylindrical dish. The weak adhesion between the gel and the lateral wall of the dish allows for the shrinkage to be unidirectional along the vertical axis and the gel to remain in contact with lateral wall of the dish. We thus focus on the vertical thinning rate of the gel. 
 
 Here we choose to work with agarose, which is a natural neutral polymer extracted from a red marine algae and composed of disaccharide units \cite{Araki:1956,Matsuhashi:1990,Djabourov:2013}. Agarose is the gelling agent of culture media used in Petri dishes \cite{Marx:2013}. These gels are subjected to drying during the incubation phase of the dish at constant temperature to monitor the potential growth of bacterial colonies. Such culture media of complex composition often contain numerous other additives, including non-gelling polysaccharides such as agaropectin, sucrose, etc. that affect the water-holding capacity of the gel. Therefore, understanding the drying kinetics of agarose gels and the influence of minute amounts of additives upon the gel thinning rate in the early stage of the drying process is of practical importance. 

Here we use interferometry as a tool to measure with high accuracy the local thinning rate $\dot z(r,t)$ of agarose gels. The drying dynamics of a gel cast in a plate exhibits three phases. During most of the drying process the gel thins at constant speed, then experiences a sudden acceleration before coming to a complete stop. Experiments conducted at different positions $r_0$ along the dish radius $r$ show that the gel thinning rate remains constant except at the very end of the drying process and follows the same overall scenario. We further demonstrate that the thinning rate $\dot z(0,t)$ measured at the center of the dish is independent of the gel thickness and diameter and as such can be used as a robust observable to compare the thinning rate of agarose gels loaded with minute amounts of additives. In particular, we compare the thinning rates of gels made either of agarose or agar, i.e., mixture of agarose and agaropectin. We show that for equal contents in agarose (larger than 0.5\%~w/w), agar gels thin 40\% slower than pure agarose gels due to the presence of agaropectin. More generally, we show that the thinning rate of agarose gels is systematically reduced by addition of minute amounts ($<0.5$\% w/w) of non-gelling polysaccharides such as glucose, dextran, guar gum or xanthan gum. Finally, we show that additives of larger molecular weights tend to decrease the gel thinning rate more efficiently than smaller polysaccharides.    

Up to now non-gelling polysaccharides such as sucrose, glucose, maltose, xanthan gum, etc. have been shown to impact the formation of agarose gels \cite{Watase:1992,Nishinari:1992,Russ:2013} and their mechanical properties \cite{Watase:1990,Nagasaka:2000,Normand:2003} when introduced in large amounts (5\% w/w or higher). An increased mass of non-gelling polysaccharides leads to ($i$) larger gel elastic modulus, ($ii$) larger strain and stress at failure and ($iii$) lower water release under external load (syneresis) \cite{Nishinari:2016}. However, the impact of non-gelling polysaccharides on the drying kinetics of agarose-based gels remains an open issue that is the topic of the present study. Furthermore, we choose here to focus on low additives content (0.5\% w/w or lower). In that range of concentrations the linear mechanical properties of agarose gel remain unaffected by the additives, which allows us to disentangle the water-holding properties induced by the additives from their impact on the gel elastic properties.   

\section{Experimental section}
\label{experimental}

\subsection{Sample preparation}
 \label{sampleprep}

 Agarose-based gels are prepared as follows: hot solutions of polysaccharides are prepared by mixing either 1\% w/w of agarose powder (CAS 9012-36-6, ref.~A9539 Sigma-Aldrich) or 1.5\% w/w of agar powder (BioM\'erieux, agarose/agaropectine 7:3, sulfate content 0.6\%  and azote content 0.45\% as determined by elemental analysis) with milli-Q water (17~M$\Omega$.cm at 25$^{\circ}$C) brought to a boil. Non-gelling polysaccharides such as glucose (CAS 50-99-7, Roquette), dextran from Leuconostoc mesenteroides (CAS 9004-54-0, Sigma Aldrich), guar gum (CAS 9000-30-0, ref.~G4129 Sigma-Aldrich), and xanthan gum (CAS 11138-66-2, ref.~G1253 Sigma-Aldrich) and which properties are summarized in table~\ref{table1} may also be added at this stage. The temperature is maintained constant at 100$^{\circ}$C for about 10~min (except for samples prepared with guar or xanthan gum which require 20~min more) and then decreased to 80$^{\circ}$C. The agar(ose) solution is prepared fresh for each series of experiment to avoid any aging associated with the agarose oxydation \cite{Whyte:1984,Mao:2016b} Gels, shaped as flat cylinders  of ($4.0\pm$0.2)~mm thick are prepared by pouring the hot polysaccharide sol in Petri dishes of 50~mm diameter made either of smooth glass [mean roughness $(0.53 \pm 0.10)$~nm as determined with a Contour Elite Bruker profilometer] or smooth polystyrene crystal (PS) (mean roughness $(11.8 \pm 3.6)$~nm) \footnote{The material the Petri dish is made of has no influence on the measurements of the gel thinning rate - see Fig.~S1 in the ESI. Nonetheless, smooth surfaces should be preferred. Indeed, rough surfaces lead to temporal fluctuations in the gel thinning rate due to the complex dynamics of the contact line between the gel and the lateral rough walls. See Fig.~S2 in the ESI and the corresponding discussion.}, and left to gelify at room temperature, i.e., T=($22\pm 2$)$^{\circ}$C. The weak adhesion of the gel to the smooth surfaces of the glass or plastic dish ensures a purely vertical shrinkage of the gel, without any sliding motion of the gel on the bottom wall of the dish, nor any detachment of the gel from the lateral wall of the dish as evidenced by the uniform and homogeneous dynamic of the local interference pattern recorded during the drying process (see below).  

\begin{table}[h]
\small
  \caption{\ Properties of the non-gelling polysaccharides used as additives in section~\ref{additives}. Note that agaropectin is naturally present with agarose in agar samples. To determine agaropectin molecular weight, we have separated the agaropectin from the agarose following the method described in ref. \cite{Hjerten:1962} and used size exclusion chromatography \cite{Mitsuiki:1999}.}
  \label{table1}
  \begin{tabular*}{0.5\textwidth}{@{\extracolsep{\fill}}lll}
    \hline
    Additives & Formula & Molecular Weight (kDa) \\
    \hline
    Glucose & C$_6$H$_{12}$O$_6$ & 0.18 \\
    Dextran & H(C$_6$H$_{10}$O$_5$)$_x$OH & 40 \\
    \textit{Agaropectin} & - & $120\pm30$ \\
    Guar Gum & - & $\sim 220$ (polydisperse) \\
    Xanthan Gum & - & $>1000$ \\
    \hline
  \end{tabular*}
\end{table}

%%%%%%%%%%%
\begin{figure}[!t]
\centering
	\includegraphics[width=\linewidth]{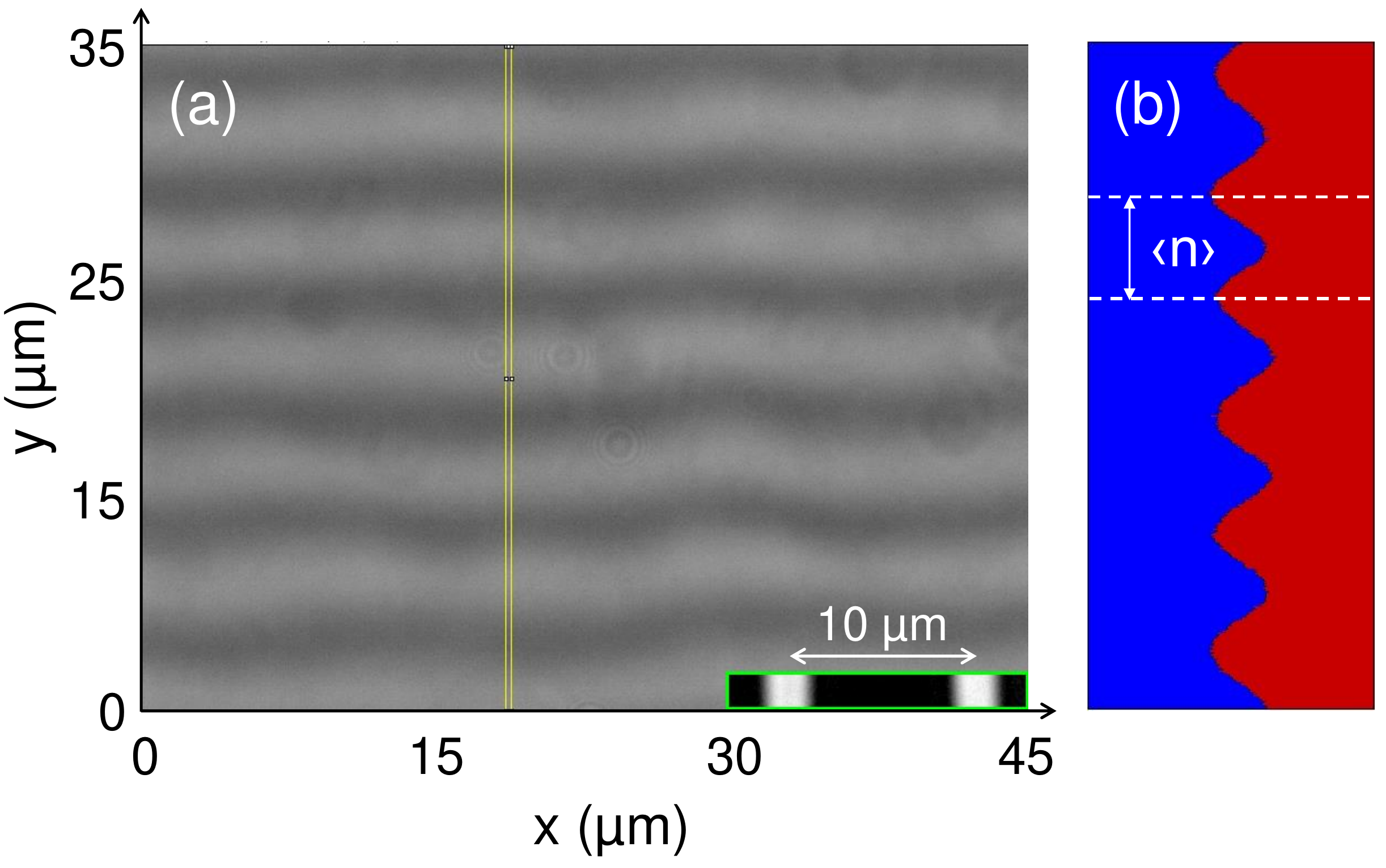}
\caption{(a) Typical instantaneous interference pattern overlayed on the image of the smooth gel surface (800~pixels $\times$ 600~pixels). Inset: image of a 10$\mu$m grid from a Leitz standard stage micrometer. (b) Projection along the $y$-axis of the gray values of all the pixels in the thin yellow vertical rectangle selection pictured in (a). The white dashed lines highlight the distance $\langle n \rangle$ between two successive dark fringes. Experiment conducted on a 1.5\% agar gel in a plastic Petri dish made of PS.     
\label{fig2}}
\end{figure} 
%%%%%%%%%%%

\subsection{Experimental setup}

A gel cast in a Petri dish as described in section~\ref{sampleprep} is left to dry at T=(25$\pm$1)$^{\circ}$C unless otherwise specified. The height variations of the gel are monitored by means of a Michelson interferometer operated in reflection mode. The principle of a Michelson Interferometer pictured in Fig.~\ref{fig1} is to split a partially coherent incident beam (red laser diode M625L3 from Thorlabs with a wavelength $\lambda=625$~nm and a coherence length of about 50~$\mu$m) into two beams: a reference beam reflected by a PZT mirror and a second beam reflected by the surface of the gel. A collimating optics produces a parallel beam splitted with a semi-transparent cube. A flat field planachromat objective without immersion (Zeiss LD Epiplan 50$\times$ with an extra long working distance of about 9.1~mm and a  numerical aperture of 0.5) forms the image of the gel in the focal plane of the camera. A second identical objective focuses the reference wave on the mirror. The mirror of the reference arm is mounted on a piezoelectric tube (Unidex 11, Aerotech) that allows us to adjust the optical phase difference between the reference and the signal arms with a nanometric accuracy so that the two waves interfere at the level of the sensitive surface of the camera (USB2 monochrome IDS uEye camera, UI124$\times$SE-M) and produce a set of almost parallel and contrasted interference fringes overlayed on the image of the smooth gel surface. A typical image recorded at the gel surface is pictured in Fig.~\ref{fig2}(a).

\subsection{Images analysis}

In order to access the thinning rate of the gel from Fig.~\ref{fig2}(a), we use the following protocol. We first adjust the mirror orientation so that the interference fringes are horizontal (or vertical) with about $5$ to $7$ fringes within the field of view (45$\mu$m x 35$\mu$m) [Fig.~\ref{fig2}(a)]. We then record a stack of 2400 TIF images during ten minutes at a frequency $f=4$~frames per second. The contrast of fringes remains almost unaltered during that time since the gel thins by about 8~$\mu$m to 15~$\mu$m, which is much less than the coherence length of the source (of about 50~$\mu$m). The average number $\langle n \rangle$ of pixels between two fringes, which corresponds to a half wavelength in terms of height differences that give rise to interferences, is determined through a projection of the gray level of all the pixels in the thin rectangular region of interest (ROI) pictured in yellow in Figure~\ref{fig2}(a) [Fig.~\ref{fig2}(b)].

\begin{figure}[!t]
\centering
	\includegraphics[width=\linewidth]{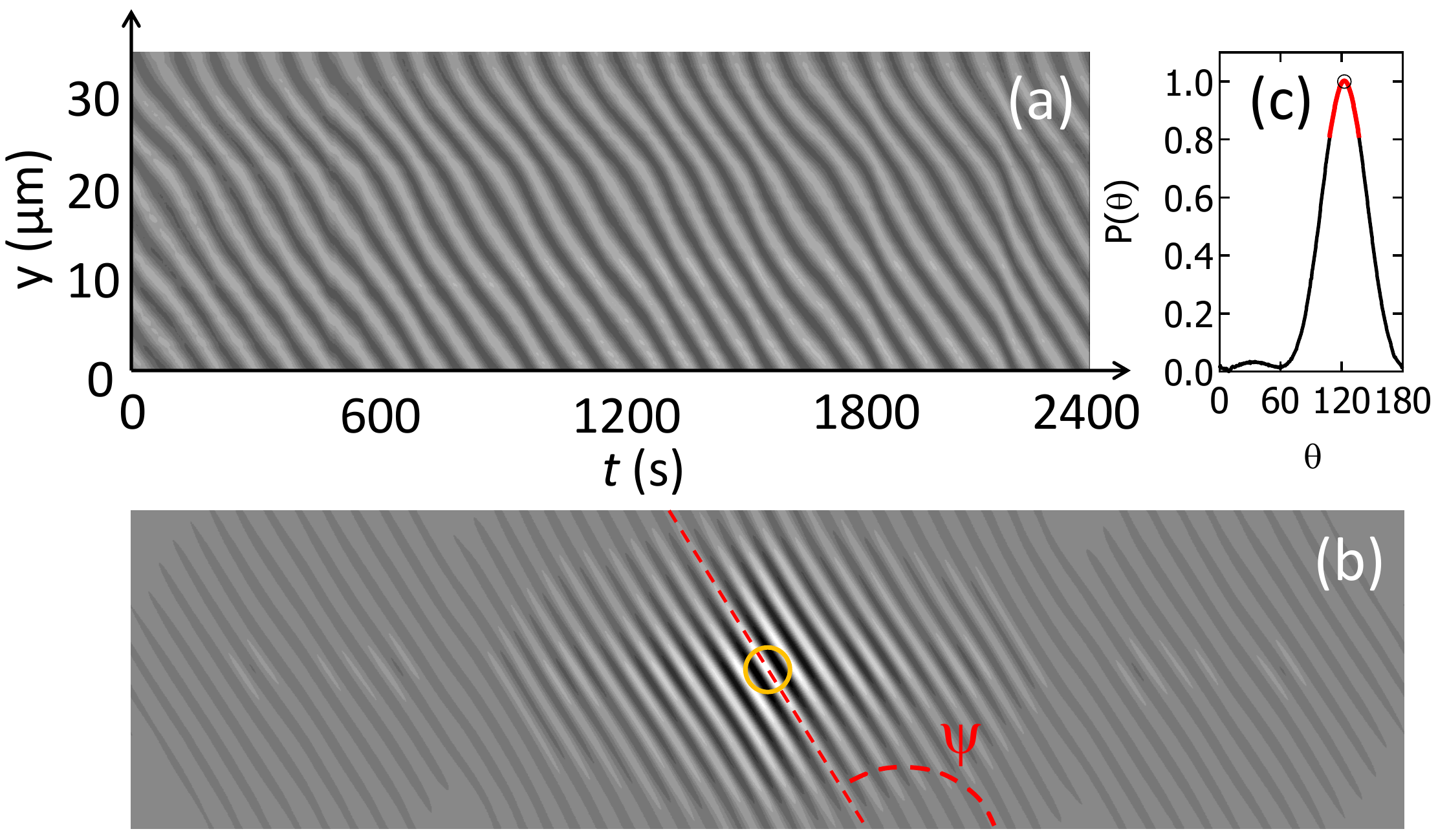}
\caption{(a) Spatiotemporal diagram $T(y,t)$ that results from projection over 20~min of the gray values of all the pixels within the thin yellow rectangle pictured in Fig.~\ref{fig2}(a) and oriented in the direction of the fringes displacement. (b) Auto-correlation function $\mathcal{A}[T(y,t)]$ of the spatiotemporal diagram $T(y,t)$ pictured in (a). (c) Angular distribution $P(\theta)$ obtained from a radial integration of $\mathcal{A}[T(y,t)]$ inside a circular ROI of 50 pixel radius that is centered on the origin of the autocorrelation image [yellow circle pictured in (b)]. The fit of the angular distribution $P(\theta)$ by a Lorentzian function over an angular sector of $30^{\circ}$ around the maximum [red curve in (c)] allows to determine the tilt angle $\Psi=121.1^{\circ}$ of the ridge line [red dashed line in (b)], and to compute the gel thinning rate $v=15.7$~nm/s averaged over 20~min. Experiment conducted at the center of a 1.5\% w/w agar gel cast in a plastic Petri dish made of PS.   
\label{fig_corr}}
\end{figure}

The subsequent analysis of the complete stack of images allows us to compute the average thinning rate $\langle \dot z(r_0,t)\rangle$ of the gel over 10~minutes (or shorter duration if need be), where $r_0$ denotes the distance between the center of the gel and the position where the beams impacts the gel and $t$ labels the time since the start of the drying experiment. We also compute the standard deviation of the thinning rate $\delta \dot z(r_0,t)$, which informs us on both temporal fluctuations in the drying dynamics and any lateral sliding motion of the gel on the bottom of the dish. The analysis is performed using a spatiotemporal filtering method based on the idea that detecting motion is equivalent to extracting an orientation inside a correlation image. The temporal projection of the gray levels of all the pixels within the thin yellow vertical ROI pictured in Fig.~\ref{fig2}(a) and oriented in the exact direction of the fringes displacement produces a spatiotemporal diagram  $T(y,t)$ featuring a pattern of oblique lines [Fig.~\ref{fig_corr}(a)]. The average tilt angle $\Psi$ of the oblique lines with respect to the time axis is directly related to the average thinning rate of the gel at the position $r_0$ through the following relation:
\begin{equation}
\langle \dot z(r_0,t)\rangle = \frac{\lambda}{2\langle n \rangle} f \tan(\Psi) \label{eq1}
\end{equation}
To determine the tilt angle $\Psi$ and deduce the average thinning rate, we compute the auto-correlation\footnote{Although the discrete 2D autocorrelation is usually computed through a discrete Fast Fourier Transform, here we use a Discrete Fast Hartley Transform (DFHT) that reduces the computer memory required and decreases the number of arithmetic operations \cite{Bracewell:1986a,Bracewell:1986b}. Prior to any discrete transform in the Hartley space, the boundaries of the spatio temporal window of size ($h,l$) are extended up to a size ($H,L$) such that $H=2^i$ and $L=2^j$, where $i$ and $j$ are integers. Extra pixels in the novel image are padded with the averaged gray level of the original image. This procedure helps reducing high-frequency noise.} $\mathcal{A}[T(y,t)]$ of the spatiotemporal diagram $T(y,t)$. The auto-correlation shows a sharp line (or ridge line) in the center with the same average orientation $\Psi$ as the oblique lines in the spatiotemporal diagram $T(y,t)$ [Fig.~\ref{fig_corr}(b)]. A radial integration of $\mathcal{A}[T(y,t)]$ inside a circular ROI centered on the origin of the autocorrelation image [radius p = 50 pixels, yellow circle in Fig.~\ref{fig_corr}(b)] provides the angular distribution $P(\theta)$, which is well fitted by a Lorentzian function [Fig.~\ref{fig_corr}(c)] and gives the tilt angle $\Psi$ of the ridge line [red dashed line in Fig.\ref{fig_corr}~(b)]. We shall emphasize that unlike traditional PIV methods, which cross correlate two consecutive interrogation regions, the spatiotemporal methods analyzes a continuous space-time window and not a discrete displacement which offers a higher accuracy to determine the thinning rate of the gel. For an optimal precision, the method yet assumes a constant brightness of both the moving fringes and the gel image. Finally, note that the interferometer and the gel are placed in a box closed by thick curtains for the whole duration of each experiment to limit as much as possible air convection, which otherwise would affect the drying process.  

%%%%%%%%%%%
%%%%%%%%%%%

%%%%%%%%%%%
\begin{figure*}[t]
\centering
	\includegraphics[width=0.8\linewidth]{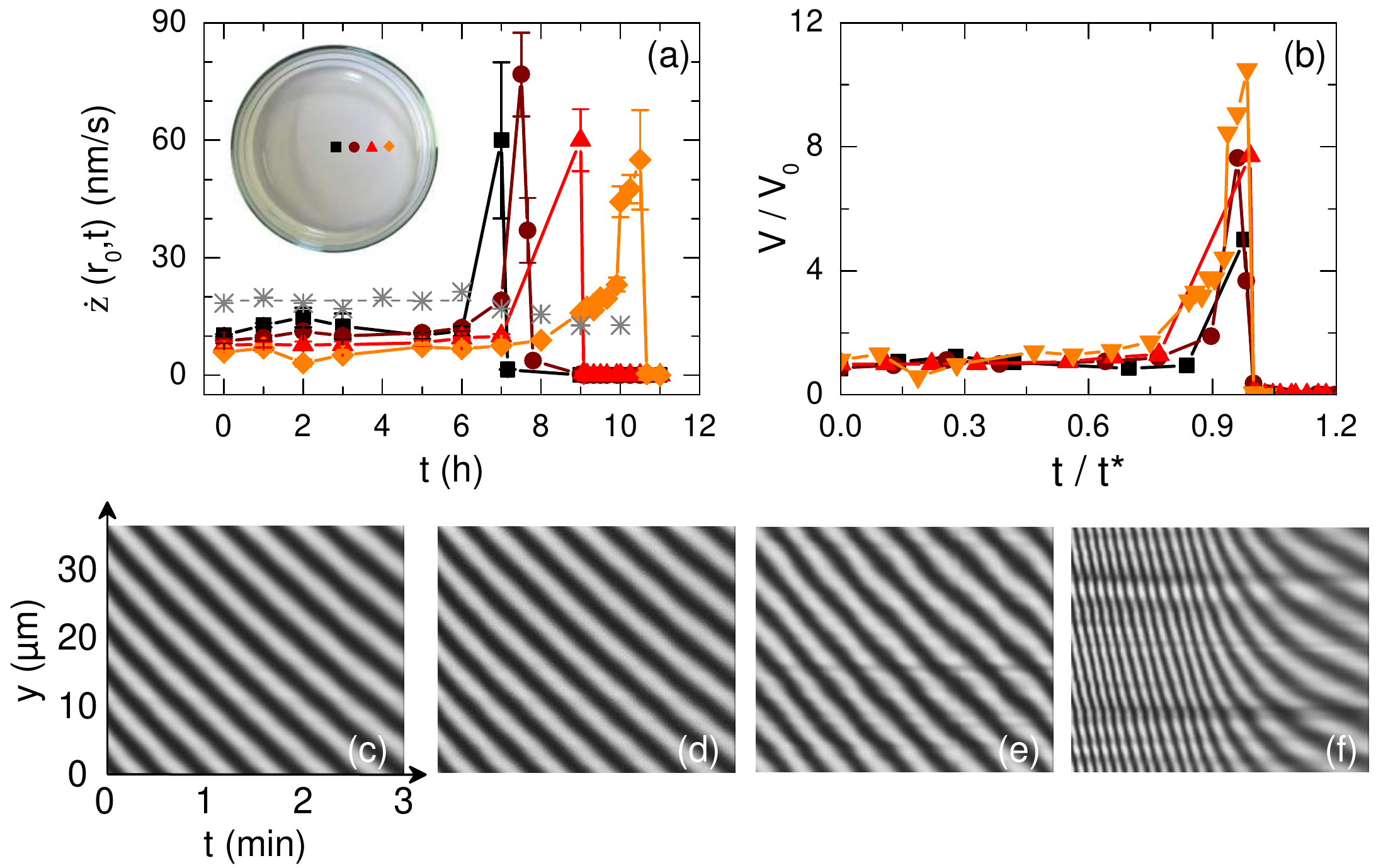}
\caption{(a) Local thinning rate $\dot z(r_0,t)$ vs time $t$ for a 1.5\% w/w agar gel ($e=1$~mm) at different locations $r_0$ from the center of the plate (diameter $2R$): $r_0/R=$0 ($\blacksquare$), 0.2 ($\bullet$), 0.4 ($\blacktriangle$) and 0.6 ($\blacklozenge$). The stars (\textasteriskcentered) stand for independent mass-loss measurements performed with a precision scale indicating an average global thinning rate of $19.1\pm 0.5$~nm/s over the first 6 hours (horizontal dashed line). (b) Same data as in (a) where the local thinning rate for each data set has been normalized by its value averaged over the first 6~hours and the time $t$ has been normalized by the time $t^*(r_0)$ associated with the largest thinning rate reached at a radial position $r_0$ during the drying process. (c)-(f) Spatio-temporal diagrams of the fringe displacement recorded over a period of 10~min during the drying process at the center of the gel and at different times $t=0$, 3, 6 and 7~h. The auto-correlation of each of these patterns provides the average thinning rate: $10.2\pm 0.8$~nm/s (c), $12.5\pm1.0$~nm/s (d), $11.3\pm0.9$~nm/s (e) and $58\pm20$~nm/s (f). These values are used to build the curve pictured as ($\blacksquare$) in (a). Experiment conducted in a Petri dish made of glass at $T=21^{\circ}$C.         
\label{fig3}}
\end{figure*} 
%%%%%%%%%%%

\section{Benchmark measurements}

\subsection{Local vs. global measurements}

\begin{figure}[b!]
\centering
	\includegraphics[width=0.9\linewidth]{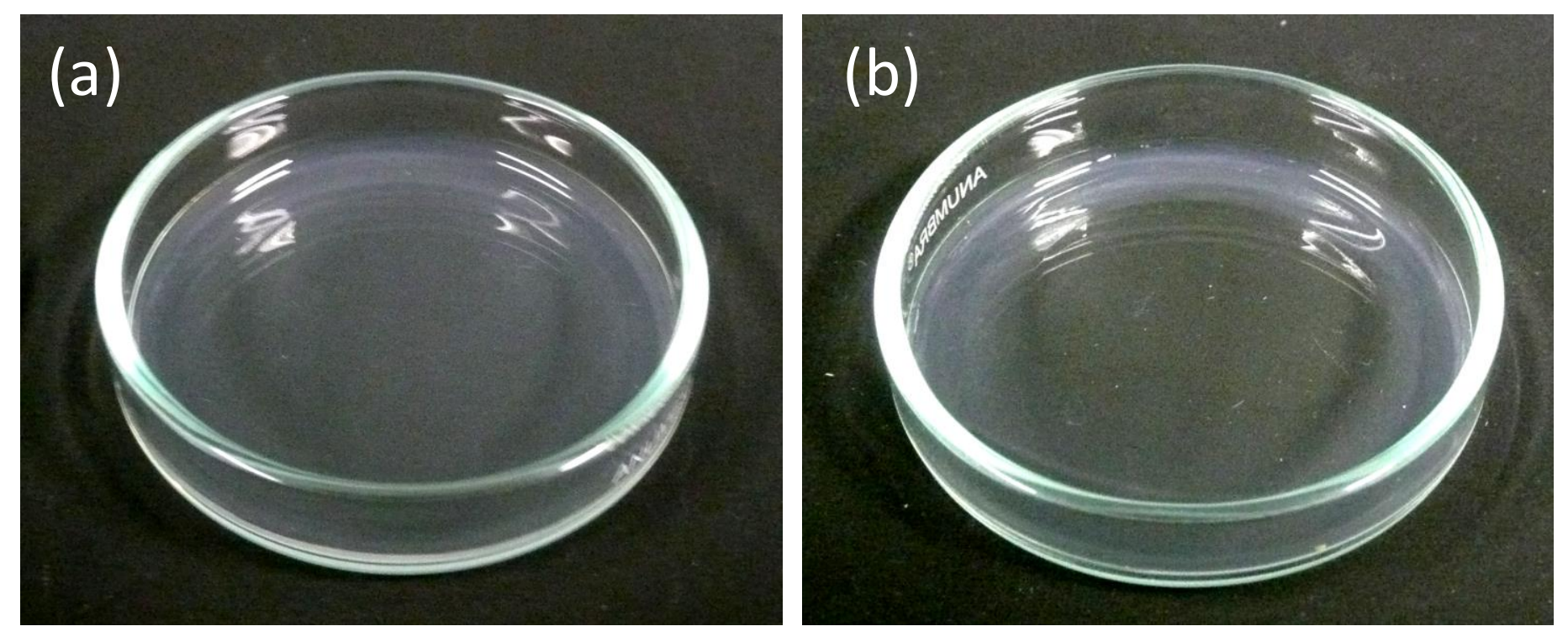}
\caption{(color online) Pictures of an agar gel cast in a Petri dish of 50~mm diameter made of glass: (a) prior to the drying experiment, and (b) 8~hours after the start of the drying experiment at T$=21^{\circ}$C. The complete drying occurs first at the center of the dish which suggests a slightly larger thinning rate of the gel at the center of the dish, as quantified in Figure~\ref{fig3}.    
\label{fig_picture}}
\end{figure}

 We first report measurements at $T=(21\pm 1)^{\circ}$C during the drying process of 1.5\% w/w agar gel of 1~mm thick and cast in a Petri dish of diameter 50~mm made of glass. The thinning rate averaged periodically over a duration of 10~min is measured as a function of time at the center of the dish and at three other locations $r_0$ along the dish radius [$r_0/R=0$, 0.2, 0.4, and 0.6 in Fig.~\ref{fig3}(a)]. During the first 6~hours, the gel thins at constant speed at the center of the dish. For $t\geq 6$~h the thinning-rate increases abruptly, goes through a maximum and stops as the drying process ends. A similar scenario is visible at the 3 other locations but shifted in time, which shows that the drying process is spatially heterogeneous: the gel thins slightly faster at the center of the dish. Indeed, the center region of the gel becomes dry first (see Fig.~\ref{fig_picture} for a picture of the gel). Despite such an inhomogeneous drying process, the evolution of $\dot z(r_0,t)$ is robust for different radial positions $r_0$ along the gel radius, as confirmed by the rescaling of the data on a single mastercurve [Fig.~\ref{fig3}(b)].
%%%%%%%%%%%
%%%%%%%%%%%

Interestingly, the local average thinning rates measured by interferometry during the first 6~hours are almost two times smaller than the global thinning rate determined by independent mass-loss measurements [(\textasteriskcentered) in Fig.~\ref{fig3}(a)]. Both local and global measurements are performed in the exact same conditions, which suggests that the water loss is more important than expected from the sole local measurements, which are limited to $r_0/R\leq 0.6$. Unfortunately, due to the lateral wall of the dish, one cannot approach the objective near the peripheral region of the sample and focus on the gel close from the wall. Direct observations of the dish from the side shows that the gel exhibits a curved meniscus of a few millimeter height in the vicinity of the wall. Water evaporates most likely faster through that meniscus due to edge effects \cite{Bocquet:2007}, which accounts for the apparent discrepancy between local and global measurements. Furthermore, the maximum thinning rate of about 60~nm/s reported in Fig.~\ref{fig3}(a) corresponds to the passage under the observation region ($r_0$)of the the triple line that marks the frontier between the dry gel, the humid gel and air. The triple line moves outwards, towards the lateral wall of the dish and water evaporation is also most likely enhanced at that very location which also accounts for the discrepancy between the local and global measurements.

%%%%%%%%%%%
\begin{figure}[!t]
\centering
	\includegraphics[width=0.95\linewidth]{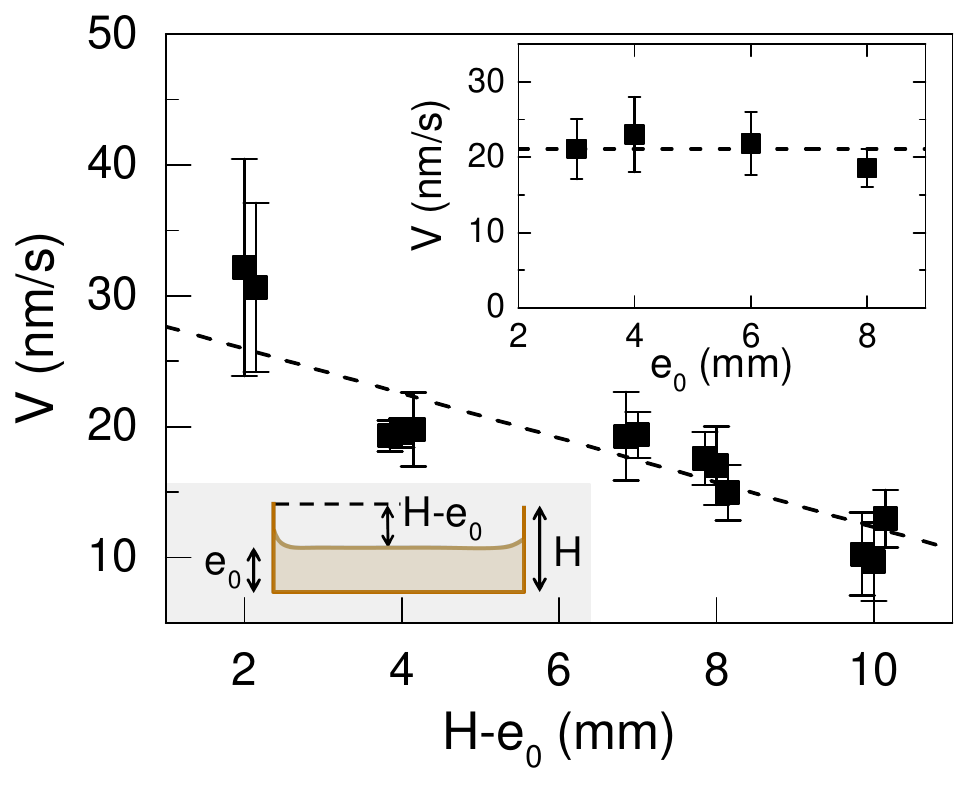}
\caption{Average thinning rate $V$ of 1.5\% w/w agar gels, measured at the center of the Petri dish vs. the height $H-e_0$ of the lateral wall relative to the gel thickness, where $H$ and $e_0$ stand respectively for the height of the lateral wall of the Petri dish and the initial thickness of the gel (see the sketch in the lower inset). The dish height is kept constant ($H=12$~mm) and the thickness of the gel is varied from $e_0=2$~mm to 10~mm, such that $H-e_0$ varies from 2~mm to 10~mm. For each value of $H-e_0$, we perform three independent experiments on gels prepared anew. Each point corresponds to an average over 10~minutes and the error bars represent the standard deviation computed over the same duration. The dashed line is the best linear fit of the data: $V=(29.3\pm2.3)-(1.7\pm0.3)(H-e_0)$. All the experiments are performed at T$=25^{\circ}$C in a plastic Petri dish made of PS, except for the experiments at $H-e_0=7$~mm which are performed in a dish made of glass. Inset: average thinning rate $V$ of 1.5\% agar gels measured at the center of the dish vs. the gel initial thickness $e_0$. The relative height of the dish with respect to the gel thickness is kept constant for all the experiments: $H-e_0=4$~mm. The dashed line stands for the average value: $V=21\pm1$~nm/s.         
\label{fig4}}
\end{figure} 
%%%%%%%%%%%

In conclusion, we shall keep in mind from this first section that the interferometer makes it possible to monitor with a high accuracy the gel thinning rate averaged over short periods of time (typically 10~min) and its temporal evolution at various locations of the gel free surface. The thinning rate is maximum at the center of the dish and constant over the first 6~hours of the drying process. In the rest of the manuscript, we use the thinning rate $V \equiv \langle \dot z(0,t)\rangle$ measured at the center of the dish and averaged over 10~minutes as a key observable independent of time to investigate the influence of the dish geometry on the drying process and quantitatively compare the thinning rate of agarose gels loaded with polysaccharides of various molecular weights (section~\ref{additives}).    

\subsection{Role of the dish geometry}
\label{geometry}

%%%%%%%%%%%
\begin{figure}[!t]
\centering
	\includegraphics[width=0.95\linewidth]{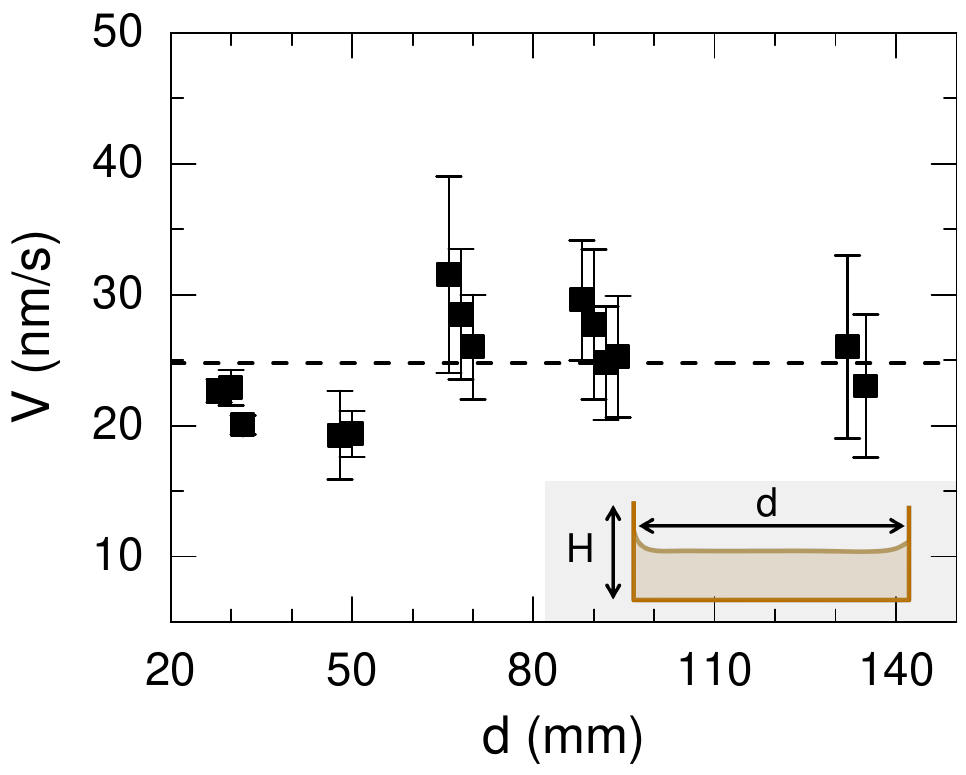}
\caption{Average thinning rate $V$ of 1.5\% agar gels measured at the center of the dish and averaged over 10~minutes vs. the dish diameter $d$. The relative height of the lateral wall compared to the gel thickness is kept constant $H-e_0=7$~mm for all the experiments. For each value of the dish diameter $d$, we perform two to four independent experiments on gels prepared anew. Error bars correspond to the standard deviation of the thinning rate over 10~minutes. The horizontal dashed line stands for the average value: $\langle V \rangle =24.7$~nm/s. Experiments conducted at T$=25^{\circ}$C in Petri dishes made of glass.  
\label{fig5}}
\end{figure} 
%%%%%%%%%%%

In this subsection, we aim at investigating the influence of the geometrical characteristics of the Petri dish on the gel average thinning rate $V$ measured at the center of the dish. First, we measure the average thinning rate of 1.5\% w/w agar gels of various initial thicknesses $e_0$ ranging from 2~mm to 10~mm, and cast in plastic Petri dishes made of PS (diameter of 50~mm and fixed height $H=12$~mm). We can thus explore the effect of the height of the dish lateral wall relative to the gel initial thickness $e_0$ measured prior to any drying. Thinning rates determined for various values of $H-e_0$ at the center of the dish are reported in Figure~\ref{fig4}. The average thinning rate of the gel $V$ decreases for increasing values of $H-e_0$, the height of the lateral wall relative to the gel thickness. A higher lateral wall relative to the gel thickness delays the water evaporation by limiting the diffusion of water vapor above the gel free surface and creating a humid area, which leads to lower thinning rates. Furthermore, we perform a series of experiments on 1.5\% w/w agar gels of various initial thicknesses $e_0$ in Petri dishes of various heights $H$, while keeping constant the height $H-e_0=4$~mm of the lateral wall relative to the gel thickness. The average thinning rate $V$ of the gel is constant within errors bars for gels thicknesses ranging between 3~mm and 8~mm, showing that for a constant value of $H-e_0$, the gel thickness does not influence the gel thinning rate [Inset in Fig.~\ref{fig4}]. The gel thinning rate at the center of the dish is thus mainly controlled by the water vapor environment located above the gel free surface.    

Second, we perform drying experiments in Petri dishes of different diameters, every other parameters kept constant ($H-e_0=7$~mm and $H=12$~mm). Thinning rates reported in figure~\ref{fig5} appear independent of the dish diameter $d$ over the following range $30\leq d \leq 135$~mm, which confirms that the average thinning rate $V$ measured at the center of the dish is a robust observable to characterize the drying of agarose gels. In the rest of the manuscript, both the dish diameter and the height of the lateral wall relative to the gel thickness are kept constant ($d=50$~mm, $H=12$~mm and $H-e_0=7$~mm) so as to investigate the influence of the agarose concentration and the presence of supplemental non-gelling polysaccharides of increasing molecular weights on the gel thinning rate.  

\section{Role of additives on the thinning-rate}
\label{additives}

\subsection{Agarose vs. agar gels}

\begin{figure}[!t]
\centering
	\includegraphics[width=0.95\linewidth]{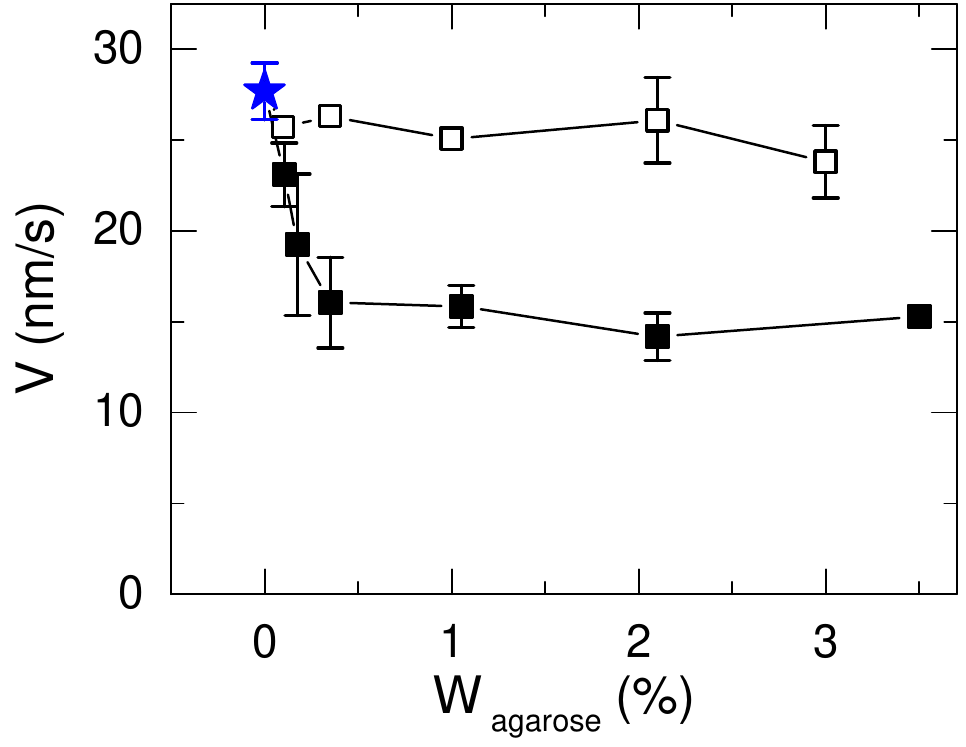}
\caption{(color online) Average thinning rate $V$ of agarose ($\square$) and agar ($\blacksquare$) gels, determined at the center of the Petri dish, vs the concentration in agarose. The agar employed here contains 70\% of agarose. The blue star (\textcolor{blue}{$\star$}) denotes the thinning rate of a water pool with the same volume as the agar(ose) gels and monitored in the exact same experimental condition (temperature $25^{\circ}$C, and dish diameter 50~mm). Error bars correspond to average standard deviation associated with three independent measurements conducted over 10~minutes each. Experiment conducted in a plastic Petri dish made of PS.       
\label{fig6}} 
\end{figure}

We now determine the average thinning rate $V$ of agarose gels of various concentrations ranging from 0.125\% to 3\% w/w. Experiments are conducted at T$=25^{\circ}$C in a plastic Petri dish (PS) of 50~mm diameter. The results reported in figure~\ref{fig6} show that the thinning rate of agarose gels is constant and independent of the agarose content up to 3\% w/w. Furthermore, the thinning rate is comparable to that of a water pool with the same initial volume as that of the gel [(\textcolor{blue}{$\star$}) in Fig.~\ref{fig6}]. This result suggests that agarose gels present negligible water-binding components and behave as passive sponges from which water evaporates freely.    

We now compare the average thinning rate $V$ of agarose and agar gels. Agar gels are composed of agarose and agaropectin, here in a ratio 7:3.\footnote{Note that the ratio depends on the agar quality and varies from one producer to another \cite{Matsuhashi:1990}. Here we have chosen to work with a standard agar commercialized by BioM\'erieux for the sake of reproducibility.} Agaropectin is a charged polysaccharide structurally similar to agarose with the same repeating units, but with higher sulfate content \cite{Duckworth:1971}. As a consequence agaropectin does not gelify. The results of drying experiments are reported in Figure.~\ref{fig6}. Over the whole range of agarose concentration explored, agar gels show systematically a smaller thinning rate than a gel that contains the same amount of agarose without any agaropectin. For concentrations in agarose lower than 0.4\% w/w, the thinning rate of agar gel decreases for increasing agarose concentrations. Above 0.4\% w/w of agarose the thinning rate is constant, independent of the agar(ose) content. This result, which holds true at lower temperature [see Fig.~S3 in the ESI] strongly suggests that agaropectin presents water-binding sites that actively slow down the water evaporation, hence reducing the gel thinning rate. 

%%%%%%%%%%%
%%%%%%%%%%%

\subsection{Other non-gelling polysaccharide additives}

\begin{figure}[!t]
\centering
	\includegraphics[width=0.95\linewidth]{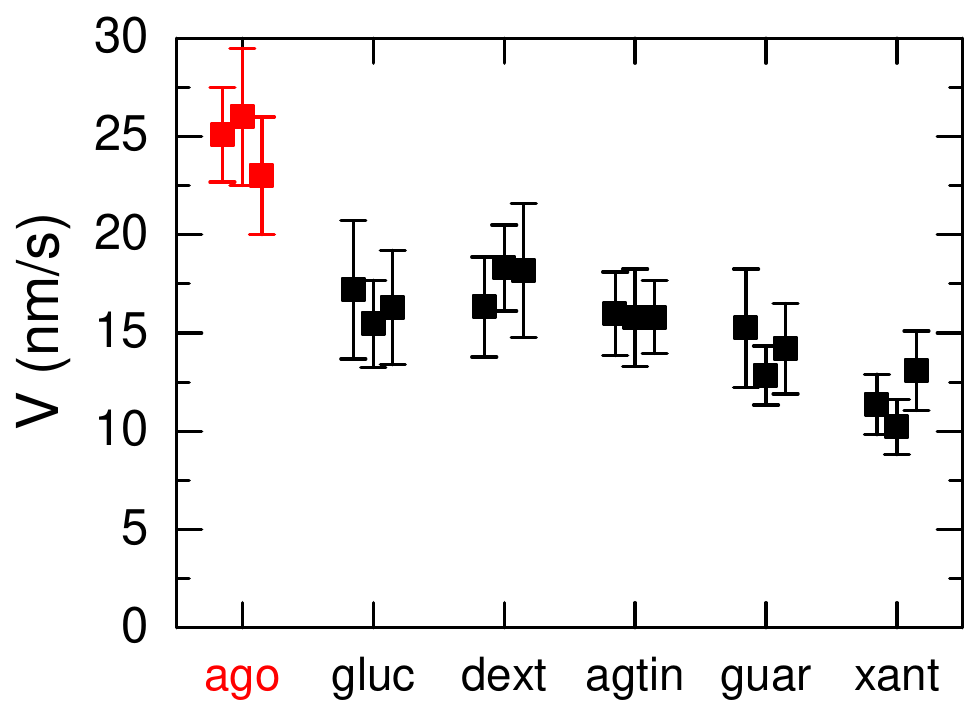}
\caption{(color online) Average thinning rate $V$ of a 1\% w/w agarose gel (``ago") and four different 1\% w/w agarose gels loaded with 0.43\% w/w of one of the following non-gelling polysaccharides: glucose (``gluc"), dextran (``dext"), agaropectin (``agtin"), guar gum (``guar") or xanthan gum (``xant"). The amount of polysaccharides is chosen so that the ratio agarose/polysaccharide is identical to the ratio agarose/agaropectin in agar, which is 7:3. For each additive three independent experiments are performed on gels prepared anew. Error bars correspond to the standard deviation of the average thinning rate over the duration of the experiment (10~minutes). Experiments conducted at T$=25^{\circ}$C in a plastic Petri dish made of PS.           
\label{fig7}}
\end{figure}

 To extend the aforementioned observations performed on agarose gels loaded with agaropectin, we have repeated the drying experiments with agarose gels loaded with other non-gelling polysaccharides of increasing molecular weights. We systematically compare the thinning rate of agarose gels loaded with 0.43\% w/w of one of the following additives: glucose, dextran, guar gum or xanthan gum to the thinning rate of a pure agarose gel determined in the exact same conditions (Fig.~\ref{fig7}). Note that such concentration corresponds to a ratio agarose/additive of 7:3 which is comparable to the ratio agarose/agaropectin in agar gels. We observe that the thinning rates of loaded gels are systematically lower than that of the pure agarose gel, and that the gel thinning rate decreases for increasing molecular weight of the non-gelling additive [see table~\ref{table1}]. This result shows that, even in minute amount, non-gelling polysaccharides play the role of water-binding sites that efficiently slow down water evaporation and delay the shrinkage of agarose gels submitted to drying.    

%%%%%%%%%%%
%%%%%%%%%%%

\section{Discussion and conclusion}

In previous work recently reviewed in ref.~\cite{Nishinari:2016}, the addition of relatively large amount of sucrose to agarose gels is reported to increase the elastic modulus $G'$ and reduce the water release, or \textit{syneresis}, which is quantified by measuring the amount of water released after an arbitrary duration (usually a few hours) from a gel submitted to an external constant load. In ref.~\cite{Nagasaka:2000}, the authors explore a large range of concentrations in sucrose, always larger than 20~\% w/w, and show a negative correlation between the value of the elastic modulus and the syneresis extent, which they interpret as a change of the gel microstructure for increasing sucrose content (i.e., a decrease in the pore size). Our study shows that the addition of non-gelling polysaccharides, even in minute amounts, is sufficient to decrease the thinning rate of the gel without any significant change in the gel elastic modulus. Indeed, independent measurements of the elastic modulus of the agarose gels charged with non-gelling polysaccharides show that the amounts of additives used here do not impact the value of the elastic modulus $G'$ (see Fig.~S4 in the ESI). Our study therefore proves that in the range of low concentrations, non-gelling polysaccharides affect the thinning rate of agarose gel without modifying their viscoelastic properties. The lower thinning rates in the presence of additives are most likely due to specific interactions between the non-gelling polysaccharides and water molecules, such as hydrogen bonds.   

To conclude, using reflexion interferometry as a local investigation tool to monitor the thinning rate of agarose gels cast in Petri dishes and left to dry at constant temperature, we have shown that the gel drying kinetics is extremely sensitive to the relative heights $H-e_0$ of the dish lateral walls with respect to the gel thickness. For a fixed value of $H-e_0$, the thinning rate of the gel measured at the center of the dish is a robust observable that can be measured precisely through the spatiotemporal filtering method introduced in the present article and used to compare the thinning rates of gels loaded with different chemical additives. While the thinning rate of agarose gel does not depend on the agarose concentration, the presence of agaropectin, or any other non-gelling polysaccharide, reduces significantly the gel thinning rate up to 40\%. 

Finally, the approach followed in the present contribution should be useful to investigate in a systematic fashion the influence of other additives such as ions, surfactants, etc. on the drying kinetics of agarose gels and more generally on a wide range of biopolymer gels.

\section*{Acknowledgments}    
This work received funding from BioM\'erieux and the ANRT under the CIFRE program, Grant Agreement No.~112972. The authors acknowledge B.~Pouligny (CNRS-CRPP) and F.~Villeval (BioM\'erieux) for stimulating discussions, J.~Martinez (BioM\'erieux) for his help with the experiments of size exclusion chromatography, as well as the MIT Writing and Communication Center for communication advice.

\clearpage 

\appendix

\onecolumngrid 

\setcounter{figure}{0}

\section{Supplemental material}

%%%%%%%%%%%
\begin{figure}[!ht]
\centering
	\includegraphics[width=0.95\linewidth]{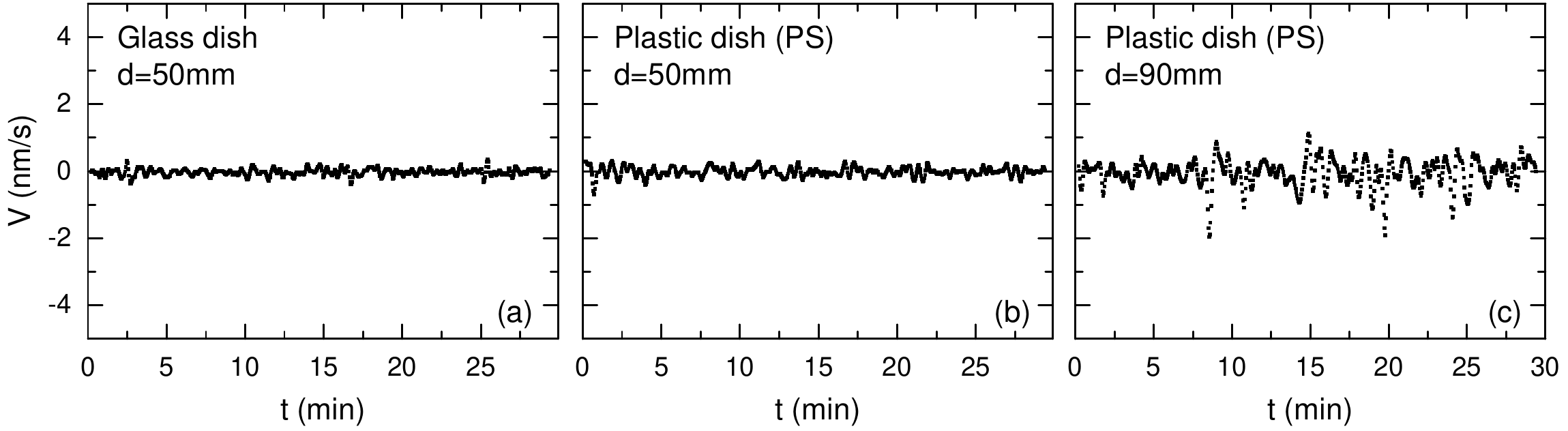}
\caption{Scrolling velocities of the interference fringes on the bottom plate of empty dishes vs time. Measurements performed at the center of the dish: (a) glass dish, 50~mm diameter; (b) plastic dish (PS), 50~mm diameter; (c) plastic dish (PS), 90~mm diameter. The root mean square of the velocity fluctuations in (a), (b) and (c) is respectively $\delta V=0.1$, 0.15 and 0.5~nm/s.        
\label{supfig1}}
\end{figure} 
%%%%%%%%%%%

\twocolumngrid

\textbf{Supplemental Fig.1} In order to determine the influence of the compliance of the Petri dish on the measurements of the gel thinning rate, we have performed a series of experiments on empty dishes, i.e. in the absence of any gel. The dish is placed empty in the arm of the interferometer. Changes in the interference pattern (recorded for 30~min at the center of the dish) now result from the thermal-induced deformations of the bottom plate. The same analysis as discussed in subsection~2.3 of the main text is applied to the interference pattern. The thermal-induced deformation of the bottom plate are quantified for a dish of 50~mm diameter either made of glass [Fig.~\ref{supfig1}(a)] or polystyrene crystal (PS) [Fig.~\ref{supfig1}(b)], and for a 90~mm diameter dish made of polystyrene crystal (PS) [Fig.~\ref{supfig1}(c)]. The two dishes of 50~mm diameter show thermal motions of comparable amplitude, whereas the more deformable plastic dish of 90~mm diameter exhibits thermal deformations of significantly larger amplitude. We have therefore chosen to work in the main text with Petri dishes of 50~mm diameter to minimize the contribution of the dish deformability to the measurements of the gel vertical shrinkage.   

%%%%%%%%%%%
\begin{figure}[b!]
\centering
	\includegraphics[width=0.95\linewidth]{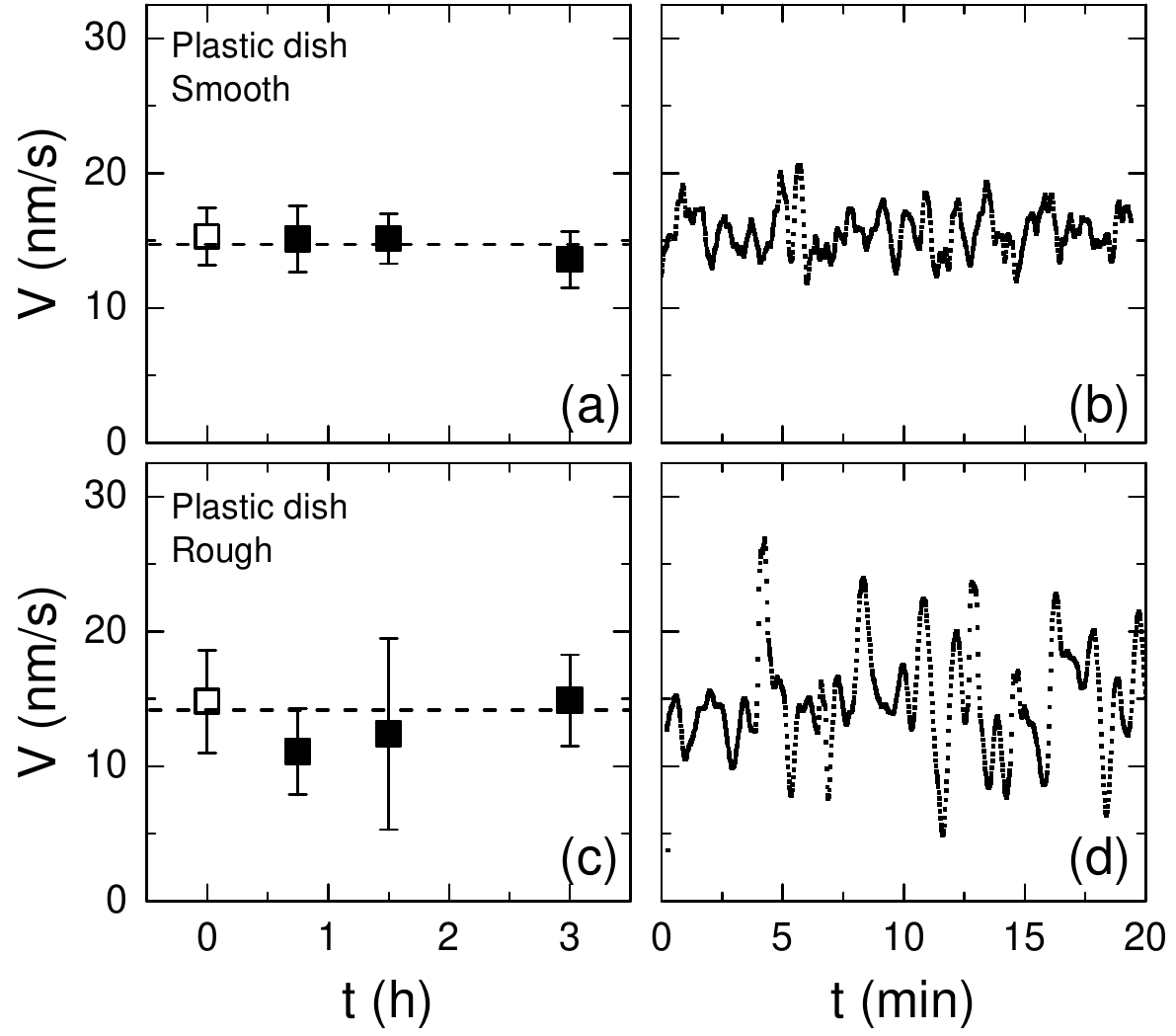}
\caption{Average thinning rate of a 1.5\% agar gel in a smooth (a) and rough (c) plastic Petri dish PS, determined at different times during a drying experiment of 3~hours. Each point corresponds to an average over 20~min. Error bars stand for the standard deviation of the average thinning rate computed over 20~min. Horizontal dashed lines in (a) and (c) stand for the average of the data. (b) [resp. (d)] shows the time-resolved evolution of the gel thinning rate for the data associated with the first point pictured as $\square$ in (a) [resp. (c)]. Experiments conducted at T=25$^{\circ}$C in a 50~mm diameter plastic Petri dish made of PS.     
\label{supfig2}}
\end{figure} 
%%%%%%%%%%%

%%%%%%%%%%%
\begin{figure}[ht!]
\centering
	\includegraphics[width=0.9\linewidth]{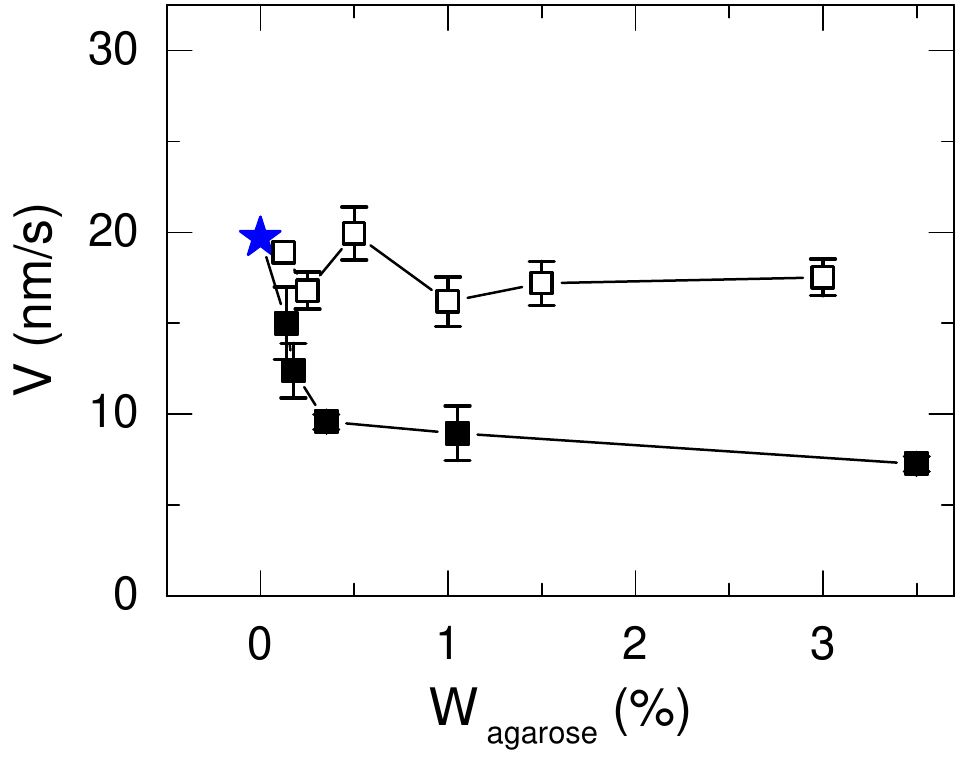}
\caption{(color online) Average thinning rate $V$ of agarose ($\square$) and agar ($\blacksquare$) gels, determined at the center of the Petri dish, vs the concentration in agarose (\% w/w). The agar employed here contains 70\% of agarose. The blue star (\textcolor{blue}{$\star$}) denotes the thinning rate of a water pool with the same volume as the agar(ose) gels and monitored in the exact same experimental conditions. Error bars correspond to average standard deviation associated with three independent measurements performed over 10~minutes each. Experiments conducted at T=$20^{\circ}$C in a plastic Petri dish made of PS.   
\label{supfig3}}
\end{figure} 
%%%%%%%%%%%

%%%%%%%%%%%
\begin{figure}[t!]
\centering
	\includegraphics[width=0.85\linewidth]{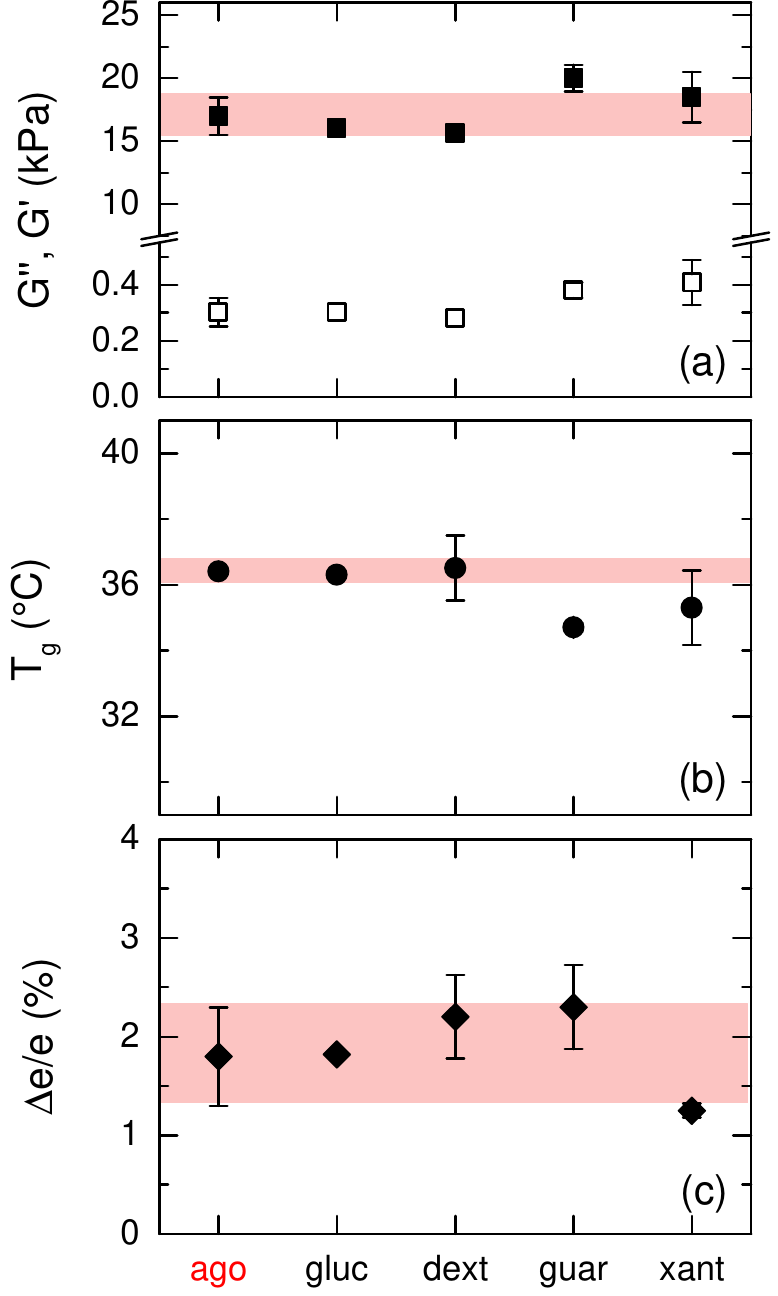}
\caption{ (a) Elastic ($\blacksquare$) and viscous ($\square$) moduli of agarose gels loaded with various non-gelling polysaccharides at 0.43\% w/w: none (``ago"), glucose (``gluc"), dextran (``dext"), guar gum (``guar")or xanthan gum (``xant"). (b) Gelation temperature for the same gels, as determined by the crossing of the elastic and viscous moduli during the sol/gel transition. (c) Relative contraction $\Delta e/e$ of the same gels during the gelation. Experiments performed at constant normal force [$F_N=(0.0\pm 0.1$)~N] in a 40~mm plate-plate geometry, with an initial gap value of $e_0=500$~$\mu$m. The red stripes highlight the range of values associated with the agarose gel without any additives.      
\label{supfig4}}
\end{figure} 
%%%%%%%%%%%

\textbf{Supplemental Fig.2} In order to quantify the impact of the surface roughness of the Petri dish on the drying dynamics of the gel, we compare two drying experiments on a 1.5\% agar gel casted in two different plastic dishes (PS), one with smooth boundary conditions [Mean surface roughness of ($11.8\pm3.6$)~nm as determined by profilometry - Fig.~\ref{supfig2}(a)], and one with rough (sand-blasted) boundary conditions [surface roughness of $(1.54\pm0.43)$~$\mu$m as determined by profilometry - Fig.~\ref{supfig2}(c)]. Over 3 hours, the average thinning rate remains constant and identical within error bars for both experiments [$\langle V_{smooth} \rangle=$(14.8$\pm$2.1)~nm/s and $\langle V_{rough} \rangle=$(14.3$\pm$4.4)~nm/s]. However, the gel thinning rate shows much larger fluctuations in the case of the rough plastic  dish (PS) [Fig.~\ref{supfig2}(d) to be compared with Fig.~\ref{supfig2}(b)]. We interpret the latter result as the  complex dynamics of the contact line between the gel, the lateral wall of the dish and the ambient air. The contact line may be trapped and suddenly released during the drying of the gel in a rough dish which generates a stick-slip like dynamics that impact the thinning rate.

\textbf{Supplemental Fig.3} Same experiment as the one reported in Figure~8 in the main text. Here the drying experiments are performed at 20$^{\circ}$C instead of 25$^{\circ}$C. Results are shifted towards lower thinning rates, but the conclusions obtained at 25$^{\circ}$C hold true at 20$^{\circ}$C.

\textbf{Supplemental Fig.4} We have determined the linear viscoelastic properties of agarose gels loaded with various non-gelling polysaccharides at 0.43\% w/w. For each sample, the gelation is induced by a temperature ramp at 1$^{\circ}$C/min from 70$^{\circ}$C to 20$^{\circ}$C in a plate-plate geometry, under constant normal force [$F_N=(0.0\pm0.1)$~N]. The gap follows the change in thickness of the sample as the latter goes through the sol/gel transition. See \cite{Mao:2016} for more details. The terminal values of the elastic and viscous moduli are independent of the additive and equal to that of the pure agarose gel [Fig.~\ref{supfig4}(a)]. In a similar fashion, the gelation temperature $T_g$ of the agarose gel, defined as the intersection of the elastic and viscous moduli during the gelation process is not affected by the presence of additives [Fig.~\ref{supfig4}(b)]. Finally, the samples shrink by about 1.7\% during gelation independently of the nature of the additive. 
          
 \clearpage         
          
%\bibliography{biblio} %your .bib file
%\bibliographystyle{aipnum4-1} %the RSC's .bst file


\end{document}